\documentclass[aps, prl, 10pt, twocolumn,superscriptaddress]{revtex4-1}

\usepackage[utf8]{inputenc}
\usepackage{epsfig}
\usepackage{graphicx}
\usepackage{float}
\usepackage{dcolumn}
\usepackage{xcolor}
\usepackage{bm}
\usepackage[normalem]{ulem}
\usepackage{amsmath,bbm}

\DeclareUnicodeCharacter{2212}{-}

\begin{document}

\title{Non-Majorana states yield nearly quantized conductance in superconductor-semiconductor nanowire devices}

\author{P. Yu}
\affiliation{University of Pittsburgh, Pittsburgh, PA 15260, USA} 
\author{J. Chen}
\affiliation{University of Pittsburgh, Pittsburgh, PA 15260, USA} 
\author{M. Gomanko}
\affiliation{University of Pittsburgh, Pittsburgh, PA 15260, USA} 
\author{G. Badawy}
\affiliation{Eindhoven University of Technology, 5600 MB, Eindhoven, The Netherlands} 
\author{E.P.A.M. Bakkers}
\affiliation{Eindhoven University of Technology, 5600 MB, Eindhoven, The Netherlands} 
\author{K. Zuo}
\affiliation{Center for Emergent Matter Science (CEMS), RIKEN, Wako 351-0198, Saitama, Japan} 
\author{V. Mourik}
\affiliation{Centre for Quantum Computation and Communication Technologies, School of Electrical Engineering and Telecommunications, UNSW Sydney, Sydney, New South Wales 2052, Australia} 
\author{S.M. Frolov}
\email{frolovsm@pitt.edu}
\affiliation{University of Pittsburgh, Pittsburgh, PA 15260, USA} 
\date{\today}

\begin{abstract}
Conductance at zero source-drain voltage bias in InSb nanowire/NbTiN superconductor devices exhibits peaks that are close to a quantized value of $2e^2/h$. 
The nearly quantized resonances evolve in the tunnel barrier strength, magnetic field and magnetic field orientation in a way consistent with Majorana zero modes. 
Our devices feature two tunnel probes on both ends of the nanowire separated by a 400 nm nanowire segment covered by the superconductor. 
We only find nearly quantized zero bias peaks localized to one end of the nanowire, while conductance dips are observed for the same parameters on the other end.  
This undermines the Majorana explanation as Majorana modes must come in pairs. 
We do identify states delocalized from end to end near zero magnetic field and at higher electron density, which is not in the basic Majorana regime. 
We lay out procedures for assessing the nonlocality of subgap wavefunctions and provide a classification of nanowire bound states based on their localization.
\end{abstract}

\maketitle

Quantized conductance, although counter-intuitive in classical physics, is confirmed in many different systems when electron transport is dominated by a few one-dimensional channels.
Metrologically robust conductance quantization is observed when backscattering of electron waves is suppressed, in edge states of quantum Hall systems \cite{klitzing1980}. 
Quantized conductance in ballistic channels is also well-established \cite{vanWees1988}, but it can be scrambled by scattering\cite{houten2005quantum}. 
Quantization in helical channels, where opposite spin carriers travel in opposite directions, has been predicted theoretically, but so far not found to be robust in experiments \cite{klinovaja2011,quay2010,nowack2013,konig2007}.
Conductance quantization has also been predicted at low transmission in Majorana transport \cite{Sengupta2001,Law2009,Flensberg2010, Wimmer2011}, and its observation is still an ongoing effort.

Majorana modes are manifestations of topological superconductivity which can be engineered, among other systems, in semiconductor nanowires by leveraging intrinsic spin-orbit coupling, placing them in contact with conventional superconductors and at the same time subjecting them to external magnetic fields \cite{Oreg2010,Lutchyn2010,Mourik2012,Albrecht2016,Deng2016}.
Majorana modes possess non-trivial properties including pinning to zero energy, quantum state delocalization and non-abelian exchange \cite{Read2000,MOORE1991,Nayak2008}.

Tunneling into a Majorana bound state would be seen as a zero-bias conductance peak with its height quantized at exactly $2e^2/h$ \cite{Sengupta2001,Law2009,Flensberg2010}. 
However, this quantization does not survive finite temperature \cite{Pientka2012,Prada2012,Lin2012,Rainis2013}, finite system size \cite{prada2012transport}, or when tunneling at zero energy is also allowed into non-Majorana states \cite{Liu2017}.
Approximate quantization may still be accessible under realistic experimental conditions when tunnel coupling dominates over both temperature and Majorana splitting.
A tunnelling rate-independent conductance is then expected at a nearly quantized level \cite{Wimmer2011}.
A recent theory however proposed that approximate conductance quantization may be achieved without Majorana modes by fine-tuning trivial low energy states \cite{Pan2020}.

Here we fabricate three-terminal InSb nanowire devices in the Majorana configuration, allowing tunneling measurements across both nanowire ends. 
On the left side, a zero-bias conductance peak (ZBCP) with nearly $2e^2/h$ absolute conductance appears at finite magnetic field and persists over a significant field range. 
This ZBCP is also persistent upon tunnel barrier tuning and only appears when the magnetic field is pointing along the nanowire,  all of which is consistent with Majorana theory. 
However, we do not observe an accompanying peak at the right end of the nanowire at the identical settings. 
Given the NbTiN superconductor-covered nanowire segment length of 400 nm, exceeding the nanowire diameter only by a factor of 4, little space is left to locate the second Majorana away from the right end. 
This rules out the possibility that the observed nearly-quantized ZBCPs are due to well-separated Majorana bound states. 
Instead, we attribute the left ZBCP to topologically trivial Andreev bound states localized on the left end of the nanowire. 
Upon careful search, we do not find correlated transport resonances simultaneously at both ends of the nanowire at finite magnetic fields sufficiently large for Majorana bound states to be observed. 
We find states delocalized between left and right ends only around zero field and at more positive gate voltages, away from the single-subband regime. They therefore constitute trivial states extending underneath the entire superconducting contact. 
We conclude that further efforts are required to understand and optimize nanowire devices, while any claim of Majorana bound state observation should be verified in the three-terminal geometry.

\begin{figure}[ht!]
\centering
  \includegraphics[width=\columnwidth]{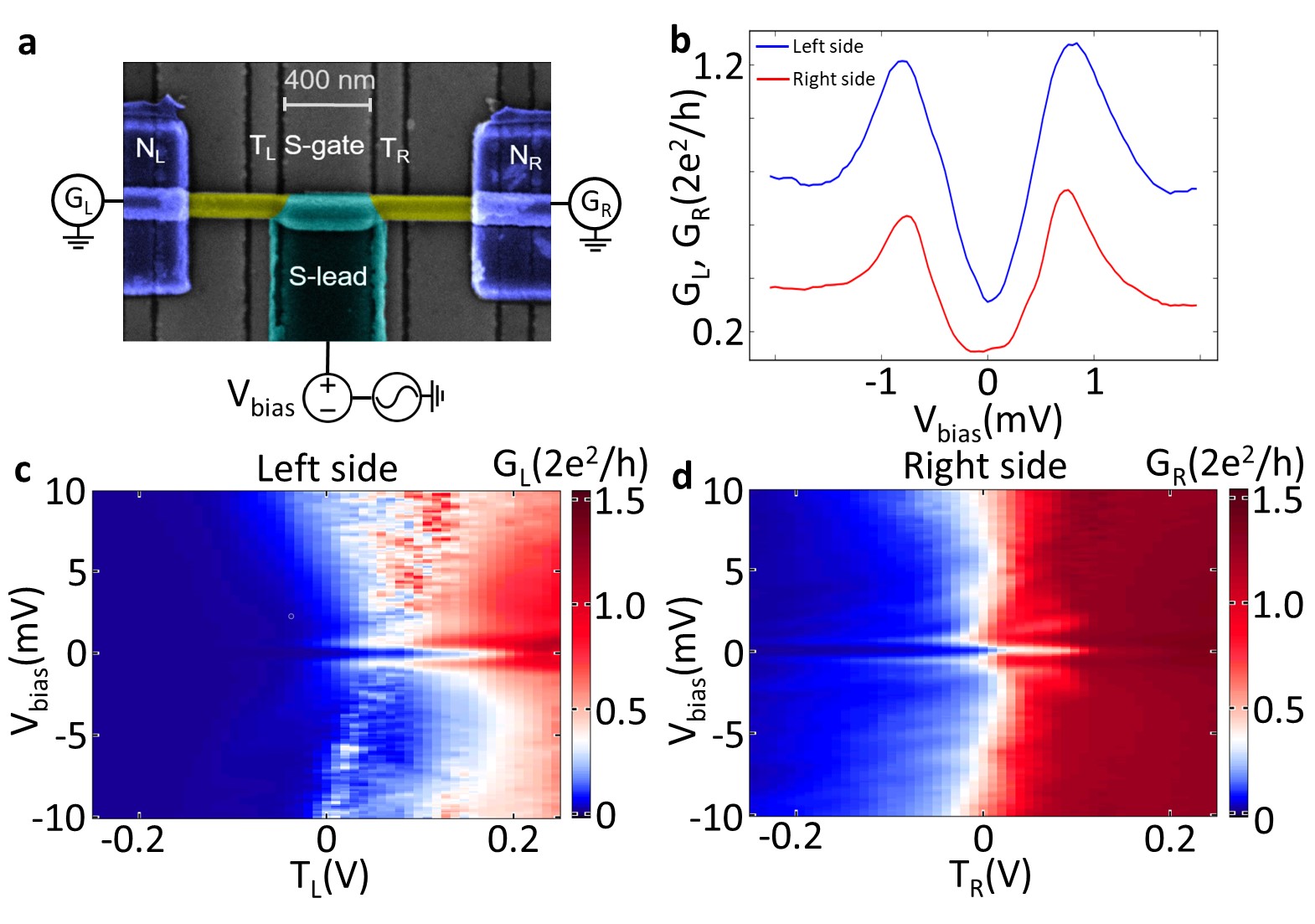}
  \caption{\textbf{Three-terminal nanowire device and basic characterizations. a}, False-color scanning electron micrograph of the measured device and the circuit diagram.
  \textbf{b}, Induced gap from both sides in the pinch-off regime where $T_L$ = -0.015 V, $T_R$ = -0.075 V and S-gate = -0.75 V. \textbf{c} and \textbf{d,} Differential conductance $G_L$ and $G_R$ as functions of tunnel-gate voltages and source-drain voltage. 
  All the other gates are set to positive voltages (open regime), and magnetic field is set to B = 0.
}
 \label{fig1}
\end{figure}

Fig. \ref{fig1}(a) shows the three-terminal hybrid device studied here. 
A metalorganic vapor-phase epitaxy-grown InSb semiconducting nanowire is covered by a NbTiN superconducting contact (S-lead) in the middle and two normal Pd contacts $N_L$ and $N_R$ at the ends. 
Beneath the nanowire, a 400nm wide electrostatic gate (S-gate) controls the electron density under the S-lead. 
We use tunnel gates $T_L$ and $T_R$ to create left and right tunneling barriers and perform tunneling spectroscopy on both sides simultaneously. 
A bias voltage is applied to the middle superconductor, while conductances $G_L$ and $G_R$ are measured at the two normal contacts with standard lock-in technique.   
Fabrication and measurement details are further described in the Methods section.

To characterize the tunnel barriers and induced superconductivity, we first perform measurements at zero magnetic field. 
Fig. \ref{fig1}(b) shows low voltage bias spectroscopy for $G_L$ and $G_R$ measured at the same setting of the S-gate. 
By setting either the left or the right tunnel gate to the pinch-off regime we observe a soft and smooth induced superconducting gaps of 760-800 $\mu$eV, typical for partially-covered NbTiN/InSb devices. 
The gap is defined by two finite bias quasiparticle peaks with no sharp resonances in between.

Figs. \ref{fig1}(c) and (d) serve for tunnel barriers characterization over a large range of voltage bias exceeding the induced gap, and conductance varying from a few percent of $2e^2/h$ to above the conductance quantum.
Both on the left and the right side, conductances evolve largely monotonically revealing no randomly formed quantum dots. 
The induced gap features persist over a wide range of conductance. 
A limited range of $T_L$ triggers charge instabilities that manifest as rapid conductance switches.
In what follows, we set the tunnel barriers to attain similar total conductance on the left and right, trying to avoid the unstable regimes.

\begin{figure*}
  \includegraphics[width=0.85\textwidth]{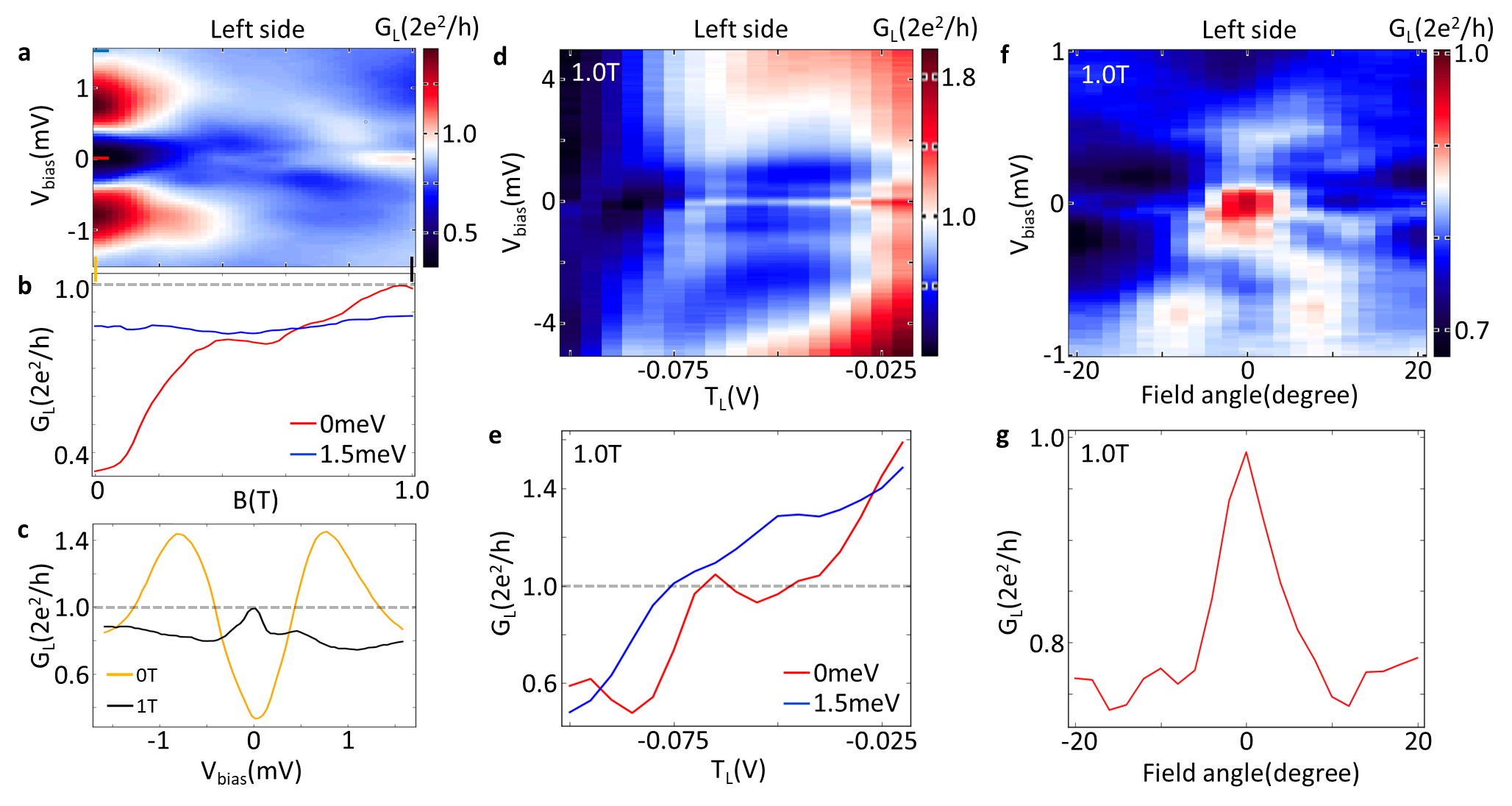}
  \caption{ \textbf{Nearly quantized zero bias conductance peak on the left side. 
  a}, Magnetic field dependence for S-gate = -0.17 V, $T_L$ = -0.045 V and $T_R$=-0.105 V. 
  The field direction is parallel to the nanowire. \textbf{b}, Linecuts taken at $V_{bias} = 0$ and 1.5 meV from \textbf{a}.
  \textbf{c}, Bias voltage linecuts from \textbf{a} at 0 T and at 1 T show the shape of the gap and the nearly quantized ZBCP.
  \textbf{d}, Tunnel-gate $T_L$ dependence of the ZBCP.
  \textbf{e}, Linecuts taken at $V_{bias} = 0$ and 1.5 meV from \textbf{d}. 
  A conductance plateau of nearly $2e^2/h$ associated with the ZBCP appear at zero bias, while the conductance above the gap evolve monotonically. 
  \textbf{f}, Field angle dependence of the ZBCP at 1 T. 
  0 degree means the field is parallel to the nanowire and perpendicular to the spin-orbit field.  
  \textbf{g,} linecut at zero bias from \textbf{f} shows a small deviation from 0 degree results in the drop of conductance from the quantized value. 
  Note a contact resistance of 4 $\mathrm{k} \Omega$ is subtracted. 
}

\label{fig2}
\end{figure*}

In Fig. \ref{fig2}, we present zero bias conductance peaks measured on the left side. 
Near zero magnetic field, $G_L$ exhibits a soft gap without resonances at low bias, as illustrated in Fig. \ref{fig1}. 
At B $>$ 0.3 T conductance resonances are observed near zero bias: they are either a ZBCP or a split peak around zero bias.
The peak conductance increases as the magnetic field increases and reaches the value of $2e^2/h$ near B = 1.0 T while the conductance beyond the gap remains nearly unchanged (Fig. \ref{fig2}(b)). 
In Fig. \ref{fig2}(c), bias voltage linecuts at 0 T and 1 T show the shape of the gap and the ZBCP. 
The ZBCP in Fig. \ref{fig2}(c) has a full width at half maximum of 150 $\mu$eV and reaches a conductance of nearly $2e^2/h$. 
The peak prominence above the background is of order 0.4*$e^2/h$. 
Without correcting for any unknown contact resistances in the device, we find a peak value of 0.8*$2e^2/h$, and to achieve exact quantization, we have to correct for 4 k$\Omega$ series resistance, a value which could be attributed to the two interface resistances.
The issue of unknown contact resistance is generic to all experiments where the exact quantization is not independently established (see supplementary materials for an extended discussion).

To study the behavior of this ZCBP against barrier transmission, we set the magnetic field to 1 T and vary the voltage on $T_L$ (Fig. \ref{fig2}(d)). 
The ZBCP only appears above $T_L = -0.07$ V and is stable for a finite range of $T_L$. 
When the ZBCP first appears, it immediately reaches its peak conductance of nearly $2e^2/h$, and maintains this conductance for a small range of $T_L$ until the ZBCP conductance increases above the quantized value predicted for Majorana modes (Fig. \ref{fig2}(e)). 
The Majorana conductance may exceed $2e^2/h$ only if the barrier has multiple transmitting channels \cite{Wimmer2011}, and indeed here the above-gap conductance reaches beyond $2e^2/h$ for more positive $T_L$.

Another check for Majorana origins of a ZBCP is in its behavior as a function of magnetic field angle with respect to the nanowire. 
Majorana states are predicted to appear only when the applied field is orthogonal to the effective spin-orbit field, previously measured to be perpendicular to the nanowire \cite{nadj2012spectroscopy, stvreda2003antisymmetric}.
Note that this measurement can be performed in NbTiN due to the large critical field but it is not practical in devices with thin Al shells because of the very small out-of-plane critical field. 
As shown in Fig. \ref{fig2}(f), the ZBCP on the left side reaches $2e^2/h$ when the magnetic field is parallel to the nanowire and perpendicular to the spin-orbit field. 
Notably, a deviation of a few degrees results in the splitting of the ZBCP and a drop in conductance. 
More magnetic field anisotropy data can be found in Fig. S11.

Overall, the behavior of the ZBCP the way it is presented in Fig. \ref{fig2} is consistent with Majorana theory, which dictates the Majorana ZBCP should emerge at finite magnetic field applied along the nanowire, persist in magnetic field, reach a peak height of $2e^2/h$, and be independent of tunnel barrier strength.

\begin{figure}[hb!]
  \centering
  \includegraphics[width=\columnwidth]{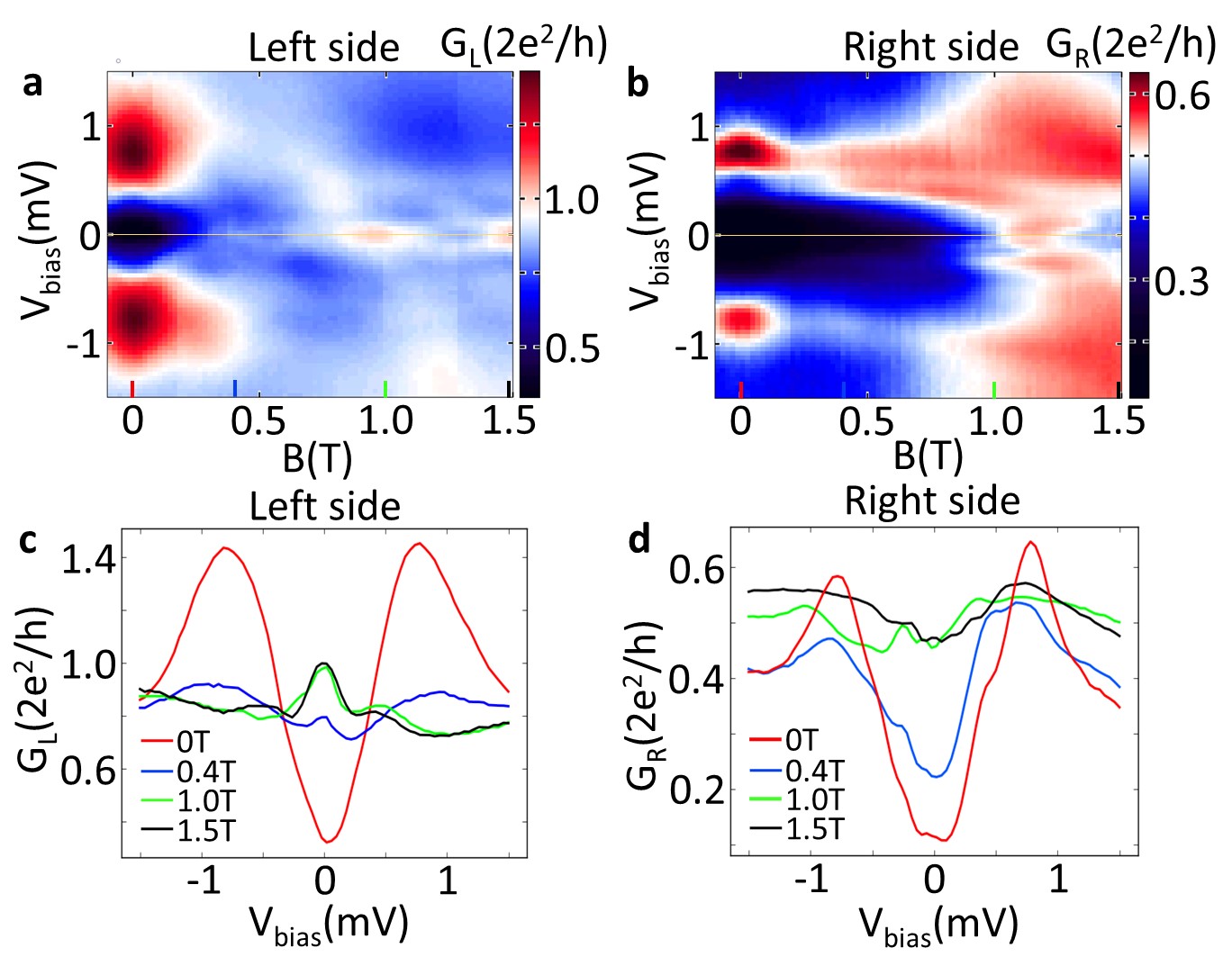}
   \caption{\textbf{Absence of zero bias peak on the right side.} \textbf{a} and \textbf{b}, Magnetic field dependence of the subgap states on the two sides from the same dataset of Fig. 2(a) now in expanded field range, where S-gate = -0.17 V, $T_L$ = -0.045 V and  $T_R$=-0.105 V. 
   A contact resistance of 4 $\mathrm{k} \Omega$ is subtracted for the left side. 
   \textbf{c} and \textbf{d}, Bias linecuts at 0 T, 0.4 T, 1.0 T and 1.5 T taken from \textbf{a} and \textbf{b} respectively.
 }
  \label{fig3}
\end{figure}

However, the simultaneous examination of conductance on both sides of the device reveals a significant deviation from the basic Majorana picture: no ZBCP is observed on the right side to accompany the left-side ZBCP from Fig. \ref{fig2}.
Fig. \ref{fig3}(a) originates from the same dataset as Fig. \ref{fig2}(a), but now we reveal a larger magnetic field range. 
Apart from the nearly quantized ZBCP described in Fig. \ref{fig2}, another peak splitting is observed at 1.2 T followed by another region of a $2e^2/h$ ZBCP around 1.5 T. 
Fig. \ref{fig3}(b) shows the simultaneously acquired conductance from the right side. 
While subgap resonances are also observed at finite fields, there is no quantized ZBCP, and in fact no ZBCP of any height is observed on the right side in this regime upon scanning $T_R$ (more information in Fig. S6). 
Linecuts taken from Fig. \ref{fig3}(a), (b) show that the left side has high conductance ZBCPs at 0.4 T, 1.0 T and 1.5 T, while the right side does not have ZBCPs at those magnetic fields.

Next, we discuss the possible interpretations of ZBCPs observed only on the left side of the device. 
A ZBCP in nanowire devices is known to have many other origins not related to Majorana modes. 
In particular, trivial Andreev bound states in quantum dots were shown to exhibit ZBCP and some degree of resonance pinning to zero bias due to gap closing and level repulsion from higher energy states \cite{LeeNatnano2014,Kells2012,Woods2019,Liu2017,Vuik2019,Moore2018,Prada2012}.
Disorder-induced ZBCPs can appear due to spectral crowding near zero bias. 
Finally, fine-tuned so-called 'class D' peaks can manifest in the intermediate mesoscopic regime which does not correspond to a large disordered ensemble of states \cite{Motrunich2001,Brouwer2011,Brouwer2011-2,Sau2013,Pikulin_2012}, this is a concept closely related to a set of a few randomly coupled quantum dots \cite{Chen2019,Woods2019}. 
There is no fundamental reason why any of such non-Majorana ZBCPs could not be tuned to have peak conductance close to a particular value, including the quantized value.

On the other hand, is it still possible that the left side ZBCP is due to a Majorana mode? In such a scenario the second Majorana bound state should be somewhere to the right of the left Majorana bound state. 
Within the relatively short S-lead segment of the nanowire, which is 400 nm in length while the nanowire itself has a diameter of 120 nm, there is little room to locate the right Majorana. 
After all, it cannot be on the right edge of the S contact where no ZBCP is observed.
It is unlikely that a possible left and right Majorana have no overlap, as both need to fit in a wire section only $\sim2$ wire diameters long. 
Majorana overlap generically induces peak splitting. 
Thus interpreting the observed ZBCP as Majorana hinges on a multitude of assumptions, each of which has to play out favorably in just the right way.

\begin{figure}[t!]
\centering
  \includegraphics[width=\columnwidth]{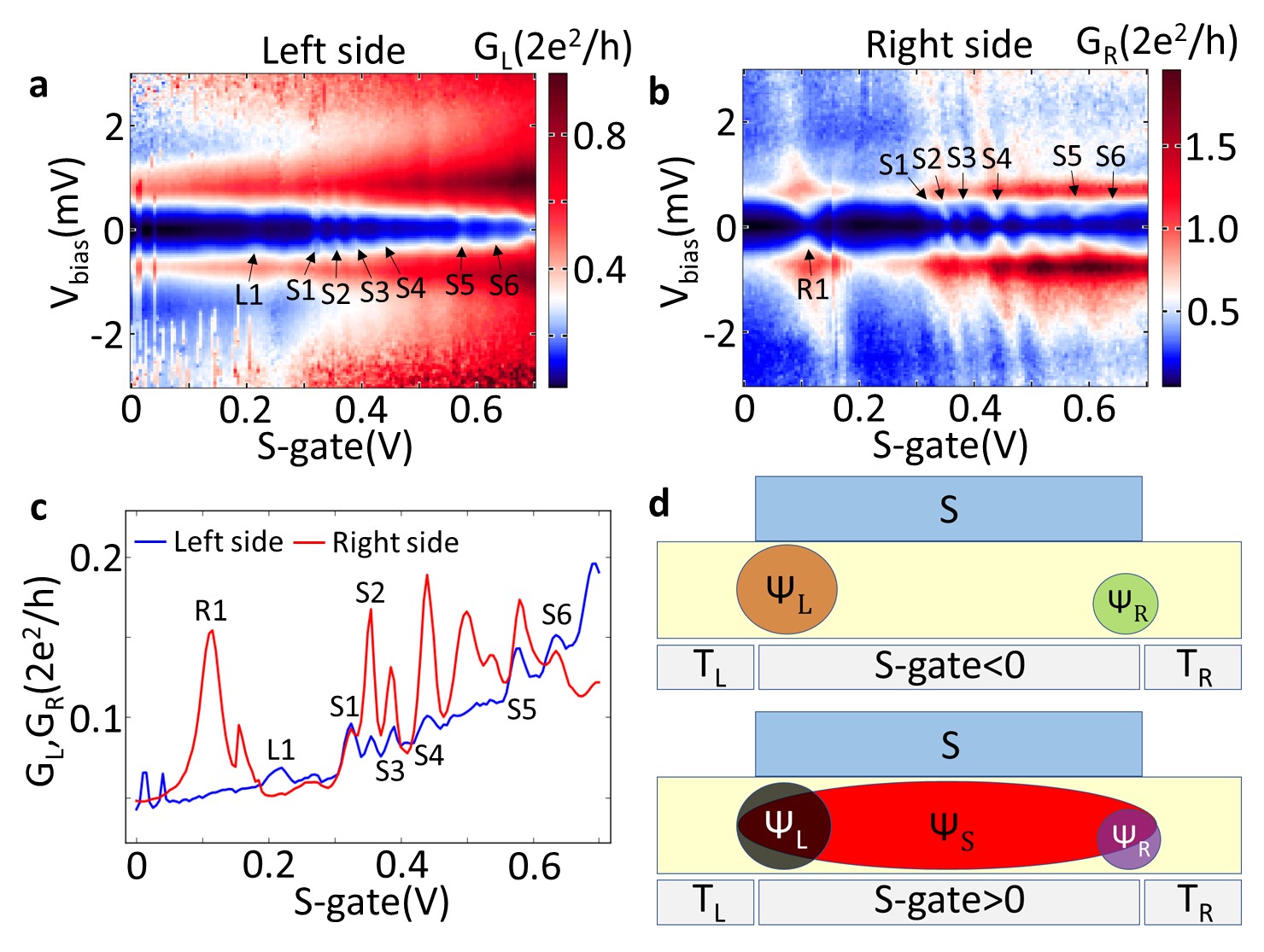}
  \caption{\textbf{Localized and delocalized states.}
  \textbf{a} and \textbf{b}, Differential conductance $G_L$ and $G_R$ as functions of source-drain voltage and S-gate voltage, showing localized and delocalized states at zero magnetic field. Tunnel gates $T_L$ and $T_R$ are set to -0.175 V and 0.09 V respectively.
  \textbf{c}, Linecuts at zero bias taken from \textbf{a} and \textbf{b} showing the overlapping delocalized states and distinct localized states. 
  \textbf{d}, Sketches of the wavefunction configuration in the device for negative and positive S-gate voltages. 
}
\label{fig4}
\end{figure}

Next we address whether any wavefunction can be confirmed as delocalized between the left and right tunnel barriers. 
This is important for demonstrating how wavefunction localization can be explored in the three-terminal geometry. 
It is also important for assessing the potential of nanowire devices in future Majorana experiments since Majorana are expected to be generated from delocalized subgap states.

In Figs. 1-3, we study subgap resonances at finite magnetic field while the S-gate is set to negative voltages, i.e. closer to the few-subband regime that is interesting for Majorana modes. 
In that regime, we do not clearly observe delocalized states. 
To find delocalized states, we ramp the magnetic field to zero and shift to S-gate$>$0, i.e. into the higher density regime (Fig. \ref{fig4}(a), (b)). 
We observe resonances that exist both above the gap as well as within the soft gap, suggesting that they are not generated by Andreev reflections.
They may be a manifestation of higher momentum wavefunctions that are localized away from the semiconductor-superconductor interface and closer to the bottom of the nanowire.
Such resonances were previously observed in similar devices \cite{Chen2017,Stanescu2016}.

Even in the positive S-gate regime we still observe resonances that are unique to either the left or the right side. 
We label such resonances with 'L' or 'R'.  
For example, when the S-gate is close to 0.11 V, there is a visible resonance on the right side, which we label 'R1'. 
On the left side, however, there is no such resonance around the same S-gate voltage. 
Instead, there is a resonance at S-gate = 0.22 V, which we label 'L1'. 
Resonances that are observed in both $G_L$ and $G_R$ simultaneously are labeled 'S'-resonances: for example six delocalized resonances S1-S6 are labeled in Fig. \ref{fig4}(a), (b), (c).
Fig. 4(d) provides a schematic summary of the findings of zero field three-terminal measurements. 
For negative S-gate voltages, two independent sets of bound states are observed on the left and the right side. 
For more positive S-gate voltages, some delocalized states are observed while the localized states are still present. 
The delocalized states only appear at low field and there is no regime with only delocalized states in this device. 
Even when S-gate$>$ 0.5 V, localized states still exist. (More data in Fig. \ref{figs4}, Fig. \ref{figs9} and in a separate forthcoming manuscript). 
We note that similar findings were reached by studying NbTiN-InSb devices with multiple S-gates in the two-terminal geometry: for more positive settings of the S-gate nearest to the tunnel gate, states sensitive to the far-away S-gates were observed (\cite{Chen2017}, supporting materials).

\begin{figure}[t!]
  \centering
  \includegraphics[width=\columnwidth]{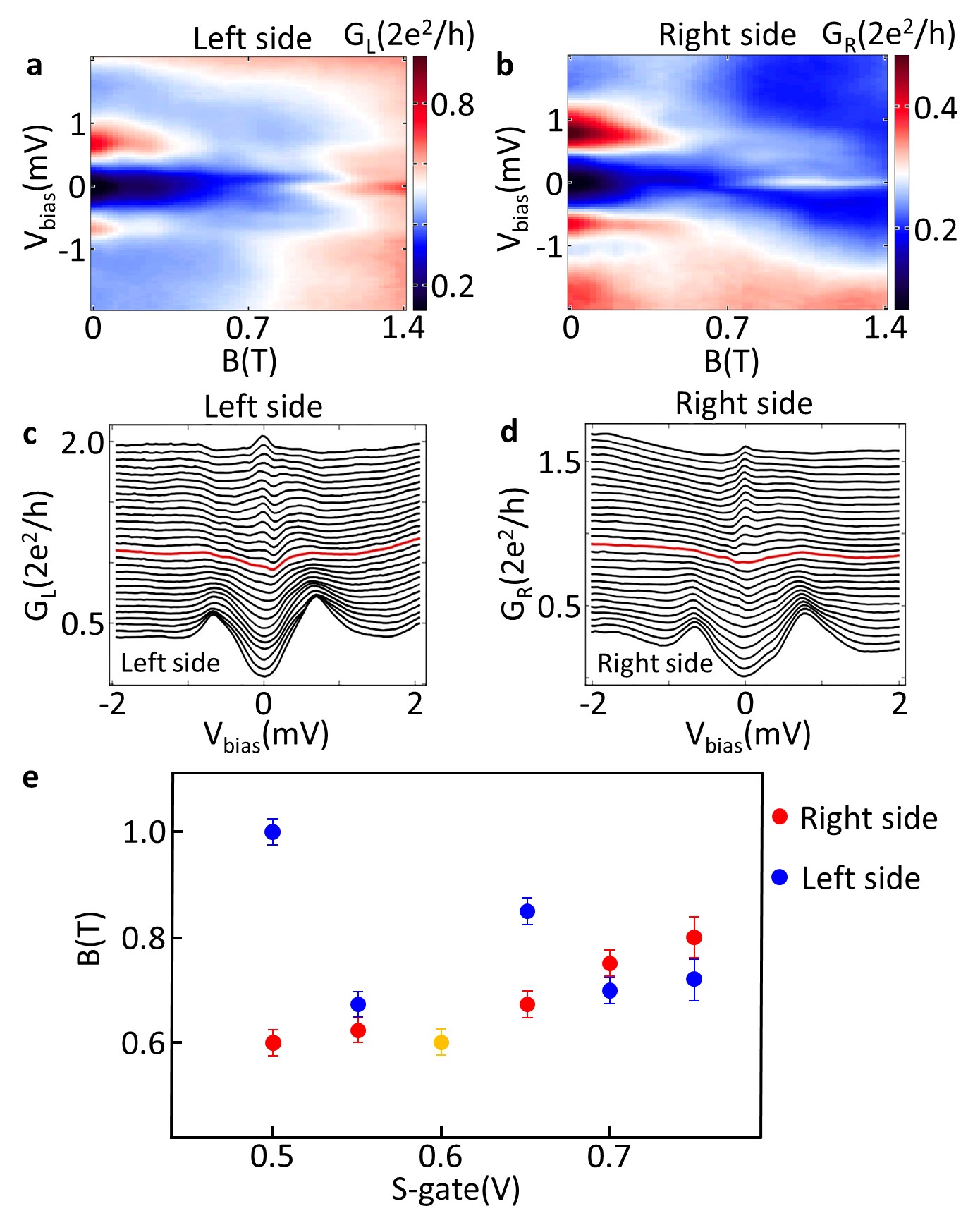}
    \caption{\textbf{Accidentally correlated ZBCPs from both sides.} 
    \textbf{a} and \textbf{b}, Magnetic field dependences of $G_L$ and $G_R$ when S-Gate = 0.6 V, $T_L$ = -0.15 V and $T_R$ = 0.09 V. 
    \textbf{c and d,} Bias linecuts taken from \textbf{c} and \textbf{d} with a interval field of 0.05 T. 
    The onset field of ZBCP (0.6 T) is highlighted in red.
    \textbf{e,} The onset magnetic fields of ZBCP for left and right sides extracted from Fig. \ref{figs10}. 
    The orange circle corresponds to data in panels \textbf{a-d}. 
    The error bars are determined by the field scan resolution.
    }
  \label{fig5}
\end{figure}

While the three-terminal geometry can be a powerful method of diagnosing the localization of wavefunctions, it is not immune to fine-tuning which can create the appearance of correlated ZBCPs (Fig. \ref{fig5}). 
For a specific setting of S-gate = 0.6 V, which is between two delocalized resonances as shown in Fig. \ref{figs9}, we observe zero-bias conductance peaks that onset at B = 0.6 T in both $G_L$ and $G_R$ and persist to over 1.4 T. 
Even though these peaks are not near the quantized conductance value, such correlation can be used to suggest that Majorana modes are observed. 
However, the correlation is not robust and vanishes when the S-gate is varied as shown in Fig. 
\ref{fig5}(e) which plots the onset fields of ZBCPs on the left and right sides as S-gate is varied. (see Fig. \ref{figs10} for full data). 
Generically, zero-bias or near zero-bias resonances exhibit no verifiable correlations between $G_L$ and $G_R$ in this device.

We emphasize that the length of the superconducting island explored in this work is 400 nm which is shorter than in many of the previous Majorana-motivated experiments. 
Since localized wavefunctions resulting in nearly-quantized ZBCPs can be observed for 400 nm long islands, there is little practical utility in exploring longer islands, e.g. 600 nm, while maintaining the same overal device fabrication method. 
Exploring shorter superconductors, such as 200 nm long islands, is likely to enhance the frequency of observing correlated subgap states in $G_L$ and $G_R$. 
However, at such short length it will be difficult to argue delocalized Majorana as the origin of the correlated signal. 
Future improvements in growth and fabrication can open the possilibty of observing delocalized Majorana wavefunctions in the three-terminal geometry. 

\subsection{Further Reading}

Background information on Majorana modes in nanowires and topological quantum computing can be found in \cite{Lutchyn2018,Stern2013}. 
For nanowire growth, details can be found in \cite{Badawy2019}. 
These are some of the previous experiments reporting Majorana modes in nanowires \cite{Mourik2012,Rokhinson2012,Deng2012,Das2012,Churchill2013} and Majorana non-locality \cite{Albrecht2016,Deng2018}.
Other three-terminal measurement works are reported in \cite{Anselmetti2019,Gramich2017,Cohen2018}. 
For theory about conductance quantization without Majorana modes, see \cite{Pan2020,Moore2018}.

\subsection{Methods}
Nanowire growth: InSb nanowires are grown in metalorganic vapour-phase epitaxy reactor using Au catalyst. 
A typical nanowire is 3-5 $\mu$m long with a diameter of 120-150 nm. 
Wire deposition: InSb nanowires are transferred under an optical microscope. 
A fine needle attached to a micromanipulator is used to pick up nanowires from the mother chip and deposit them onto the desired fine gates with sub-micrometer accuracy.
Contact deposition:  After the nanowire transfer, contact patterns are written on PMMA 950A4 resist using electron beam lithography. 
5 nm NbTi and 60 nm NbTiN are sputtered onto the nanowire with an angle of 60 degree regarding the chip substrate. 
In a separate lithography cycle, 100 nm Pd is evaporated onto the chip as normal contacts. 
Before each metal deposition, we use a standard Sulfur passivation process to remove the native oxide on the nanowire, followed by a 10 second low power in-situ argon sputter cleaning. 

Measurements are performed in a dilution refrigerator at a base temperature of 40 mK. 
Two lock-in amplifiers are used to measure the two sides of the device simultaneously using the standard low-frequency lock-in technique (77.77 Hz, 5 $\mu$V). 
Multiple stages of filtering are used to enhance signal-to-noise ratio. 
For all the measurements, a bias voltage is applied to the superconducting contact and the two normal contacts are grounded. 
We normalized the differential conductance directly measured with the two lock-in amplifiers, as described in the Appendix.

\subsection{Volume and Duration of Study}

To study the delocalized states and quantized ZBCP, 15 chips were fabricated and cooled down, on which 41 three-terminal devices were measured. 
About half of the devices had high contact resistance and were not studied in detail. 
19 devices were studied in detail, among which four devices showed delocalized states. 
Two devices showed ZBCPs with conductance close to $2 e^{2} / h$, though high conductance ZBCPs were not the focus of all cooldowns. 
For the device studied in this paper, about 7000 useful datasets were obtained within three months.

\subsection{Author Contributions}

G.B. and E.B. provided the nanowires. P.Y. and J.C. fabricated the devices.
P.Y. and M.G. performed the measurements. 
P.Y., K.Z., V.M. and S.F. analyzed the results and wrote the manuscript with contributions from all of the authors.

\subsection{Acknowledgements}
We thank S. Gazibegovic for assistance in growing nanowires. 
Work supported by NSF PIRE-1743717, NSF DMR-1906325, ONR and ARO.
\bibliographystyle{naturemag}

\bibliography{Ref.bib}

\clearpage

\newpage

\setcounter{figure}{0}
\setcounter{equation}{0}
\setcounter{section}{0}
\begin{widetext}

\section{Appendix: Supplemental Materials}

\bigskip

\renewcommand{\thefigure}{S\arabic{figure}}

\textbf{Conductance normalization procedure.} 
To obtain the actual differential conductance $d I / d V$ on both sides, we must correct the measured differential conductance from the lockin removing the contribution from the measurement circuit. 
First, we need to subtract the series resistances of the wiring, the RC filters and the input impedance of the amplifier. 
As indicated in Fig. \ref{figs1}(b), the AC lockin voltage $\Delta V_{\mathrm{ac}}$ will drop across the $\mathrm{R}_{\text {wire }}$ and the $\mathrm{R}_{\text {filters }}$ of the voltage bias line. 
This voltage drop is $\Delta V_{\mathrm{s}}=\left(\Delta I_{L}+\Delta I_{R}\right)\left(R_{\text {wire }}+R_{\text {filters }}\right)$ where $\Delta I_{L}$ and $\Delta I_{R}$ are the AC current passing through the left side and the right side of the device respectively. 
For the left side, the AC voltage drops across the second set of RC filters of the normal contact line, the DC wires and the impedance of the amplifier. 
This voltage drop is given by $\Delta V_{\mathrm{n}}=\Delta I_{L}\left(R_{\text {wire }}+R_{\text {filters }}+R_{\mathrm{im}}\right)$. 

Then the AC voltage applied to the nanowire device on the left side is $\Delta V_{L}=\Delta V_{\mathrm{ac}}-\Delta V_{\mathrm{s}}-\Delta V_{\mathrm{n}}$ and the corrected $G_{L}^{\prime}$ for the left side is:
$G_{L}^{\prime}=\frac{d I_{L}}{d V_{\mathrm{L}}}=\frac{d I_{L}}{d V_{\mathrm{ac}}} \frac{d V_{\mathrm{ac}}}{d V_{L}}=G_{L}\left(\frac{1}{1-\frac{\Delta V_{\mathrm{S}}}{\Delta V_{\mathrm{ac}}}-\frac{\Delta V_{\mathrm{n}}}{\Delta V_{\mathrm{ac}}}}\right)=G_{L}\left(\frac{1}{1-\left(G_{L}+G_{R}\right)\left(R_{\text {wire }}+ R_{\text {filters }}\right)-G_{L}\left(R_{\text {wire }}+R_{\text {filters }}+R_{\mathrm{im}}\right)}\right)$.
$G_{L}=\frac{d I_{L}}{d V_{\mathrm{ac}}}$ and $G_{R}=\frac{d I_{R}}{d V_{\mathrm{ac}}}$ are the measured differential conductances from the lock-ins connected to the left side and the right side respectively. 
Similarly, 
$G_{R}^{\prime}=G_{R}\left(\frac{1}{1-\left(G_{L}+G_{R}\right)\left(R_{\text {wire }}+ R_{\text {filters }}\right)-G_{R}\left(R_{\text {wire }}+R_{\text {filters }}+R_{\mathrm{im}}\right)}\right)$.

A unique difficulty in unambiguously establishing quantized Majorana conductance is methodological in nature.
Quantized Majorana conductance is predicted to occur even when the tunnel barrier transmission is non-monotonic and does not exhibit quantized conductance pleateaus \cite{Wimmer2011}. 
However, under realistic conditions in nanowire devices featuring some degree of disorder it is more likely that only a single quantized Majorana appears, but no higher quantized plateaus. 
As a consequence, the absolute value of conductance cannot be calibrated accurately, as only a single value of known conductance (i.e. the presumed quantized Majorana conductance) is present, which needs to be corrected for series resistances inevitably present in the measurement circuitry and the device. 
The series resistances caused by metal-semiconductor interfaces can only be roughly estimated, but not measured independently.
This methodological challenge hurdles accurately verifying the exact conductance at a supposed quantized plateau due to a Majorana.

In the present work we do consider the influence of the contact resistance. 
In Fig. \ref{fig2} and Fig. \ref{fig3}, $R_{L^{\prime}}=4\ \mathrm{k} \Omega$ was subtracted to fit the ZBCP to $2 e^{2} / h$.
This contact resistance value is also consistent with the saturation resistance determined in Fig. \ref{figs3}(a).  
It is easy to decide the contact resistance in experiments such as quantum point contact measurements because it is possible to fit the conductance plateau sequence to the anticipated quantized values. 
However, in our experiment the actual $R_{L^{\prime}}$ is hard to determine, given the fact that we don’t see a well defined conductance plateau from subband-resolved transport and the origin of the ZBCP is unlikely due to Majorana. 
Series resistance can be extracted from the saturation conductance at positive gate voltages, but this method is also prone to inaccuracy due to gate screening and other factors. 
In Fig. \ref{figs2}(c), we present a series of linecuts with different $R_{L^{\prime}}$ subtracted from the same data set of Fig. \ref{fig2}(d). 
As we can see, with $R_{L^{\prime}}$ increased from 0 to 5 $\mathrm{k} \Omega$, the height of the plateau grew from 0.8*$2 e^{2} / h$ to around 1.1*$2 e^{2} / h$. 

\clearpage

\begin{figure}[t!]
\centering
  \includegraphics[scale=0.35]{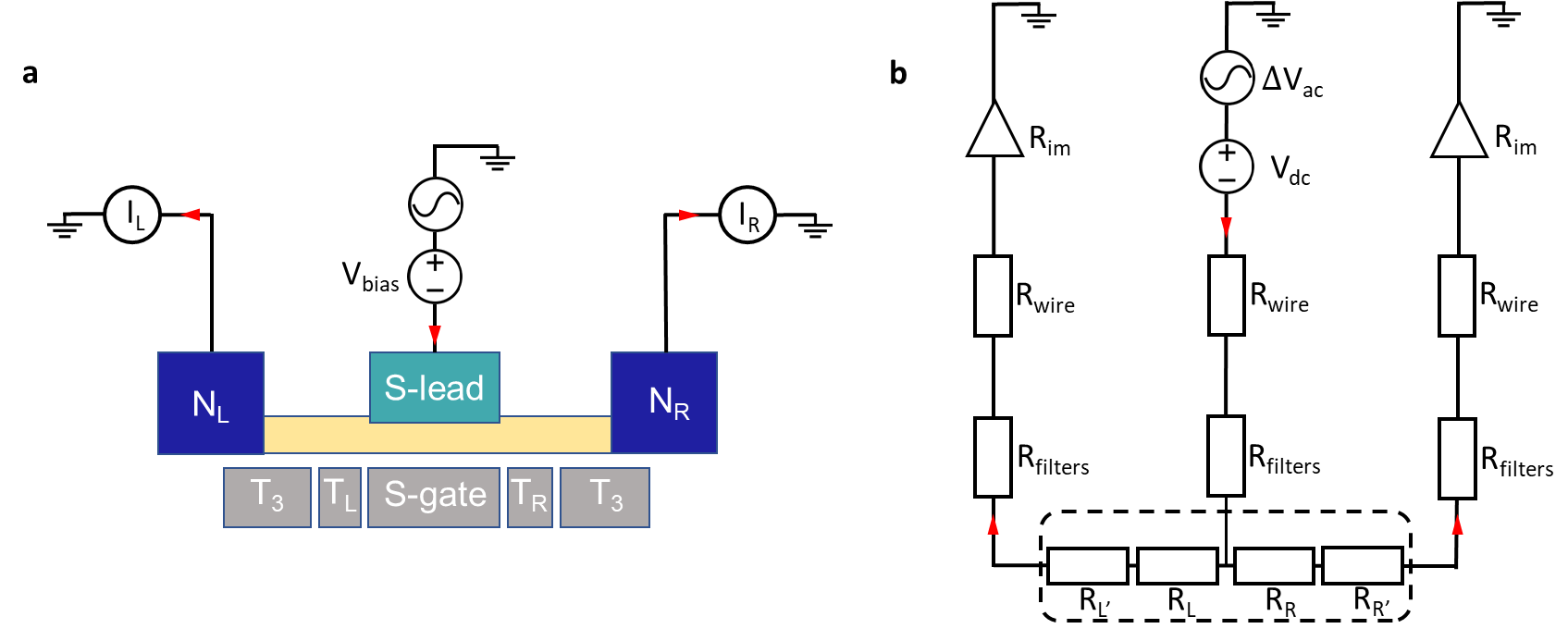}
  \caption{\textbf{Three terminal measurement setup.} 
  \textbf{a}, Schematics of the device and measurement setups. 
  Red arrows indicate the direction of dc current flow for positive bias. 
  The source-drain voltage is applied through the superconducting contact, current and differential conductance are measured simultaneously at two normal contacts. 
  The two wider tunnel gates are connected together as $T_3$.
  \textbf{b,} Simplified measurement circuit diagram representing all elements of the circuit as resistors. 
  $\mathrm{R}_{\text {filters }}$ is the resistance of RC filters and  $\mathrm{R}_{\text {im }}$ is the input impedance of the current amplifier. 
  Resistances within the dashed box are on chip. 
  They are left and right nanowire segment resistances, $R_L$ and $R_R$, as well as superconductor-semiconductor and superconductor-normal metal contact resistances, which are indicated by $R_{L^{\prime}}$ and $R_{R^{\prime}}$. 
  The exact values of contact resistances are unknown, but they can be estimated from saturation current at positive gate voltages. 
}
 \label{figs1}
\end{figure} 

\begin{figure}[b!]
\centering
  \includegraphics[scale=0.35]{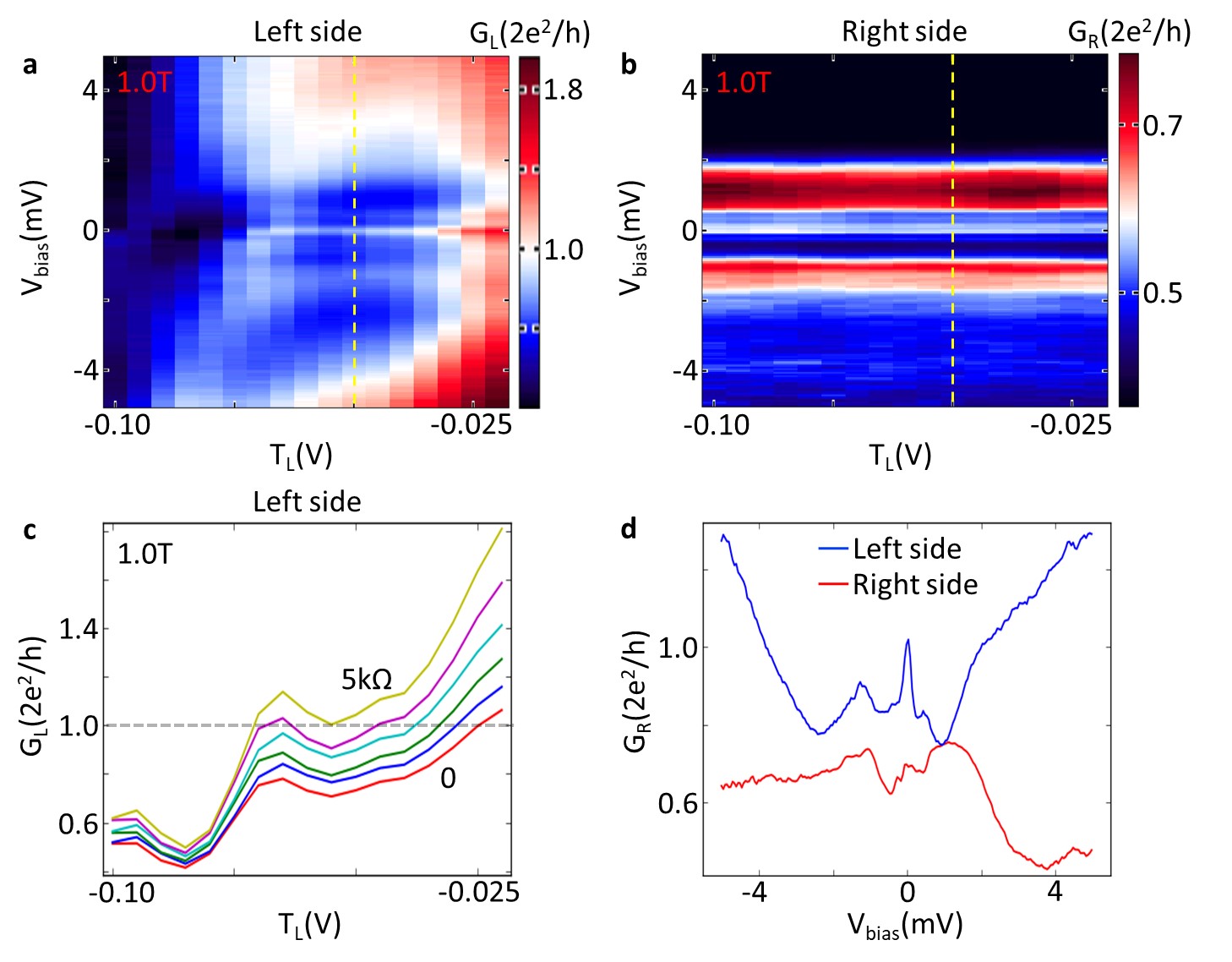}
  \caption{\textbf{Extended data of the tunnel barrier gate dependence of the ZBCP and the effect of contact resistance.} 
  \textbf{a} and \textbf{b}, Differential conductance $G_L$ and $G_R$ as functions of $T_L$ voltage and source-drain voltage from the same dataset as Fig. \ref{fig2}(d). The gate settings are S-gate = -0.17 V and $T_R$ = -0.105 V. 
  \textbf{c,} Zero bias linecuts from panel \textbf{a} with different $R_{L^{\prime}}$ subtracted. 
  When 0, 1 $\mathrm{k} \Omega$, 2 $\mathrm{k} \Omega$, 3 $\mathrm{k} \Omega$, 4 $\mathrm{k} \Omega$, 5 $\mathrm{k} \Omega$ are subtracted (from bottom to top), the conductance plateau increases from 0.8*$2 e^{2} / h$ to 1.1*$2 e^{2} / h$. 
  \textbf{d,} Bias linecuts at $T_L$ = -0.05 V (yellow dashed line) from panel \textbf{a} and panel \textbf{b} show the shape of the ZBCP on the left side and no clear ZBCP on the right side.    
}
 \label{figs2}
\end{figure}
\clearpage

\begin{figure}[t!]
\centering
  \includegraphics[scale=0.3]{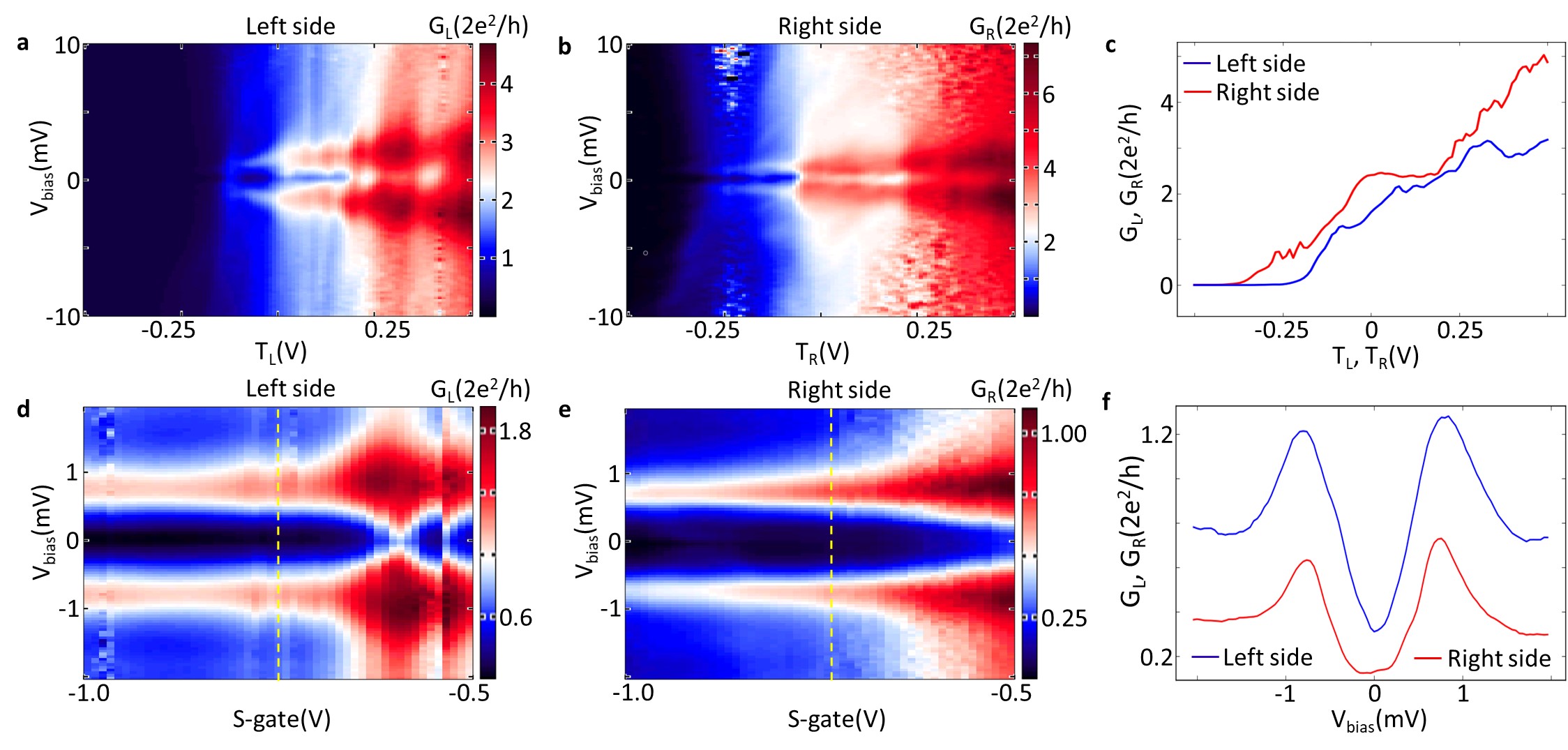}
  \caption{\textbf{Additional zero field tunnel barrier data and induced gaps.} 
  \textbf{a} and \textbf{b}, Barrier gate scans from the left and right sides respectively, while S-gate is set to 1 V. 
  The two sides show similar barrier gate dependence and overall transparency. 
  Note the left side reaches 3*$2 e^{2} / h$ at saturated regime, indicating a possible contact resistance of 3-4 $\mathrm{k} \Omega$. 
  \textbf{c,} Pinch off traces at $V_{bias}$ = 10 meV from \textbf{a} and \textbf{b}. 
  \textbf{d} and \textbf{e}, Differential conductance $G_L$ and $G_R$ as functions of S-gate voltage and source-drain voltage, while $T_L$ = -0.015 V and $T_R$ = -0.075 V. 
  While the barrier gates are set near the pinch off regime. The two sides have very similar induced gaps.
  \textbf{f}, Bias linecuts at S-gate = -0.75 $V$(yellow dashed lines) from \textbf{d} and \textbf{e}. 
  The left side has a gap with $\Delta$ = 800 $\mu$eV and the gap on the right side is about 760 $\mu$eV. 
  The error bar for the gap estimation is 40 $\mu$eV.
}
 \label{figs3}
\end{figure}

\begin{figure}[t!]
\centering
  \includegraphics[scale=0.35]{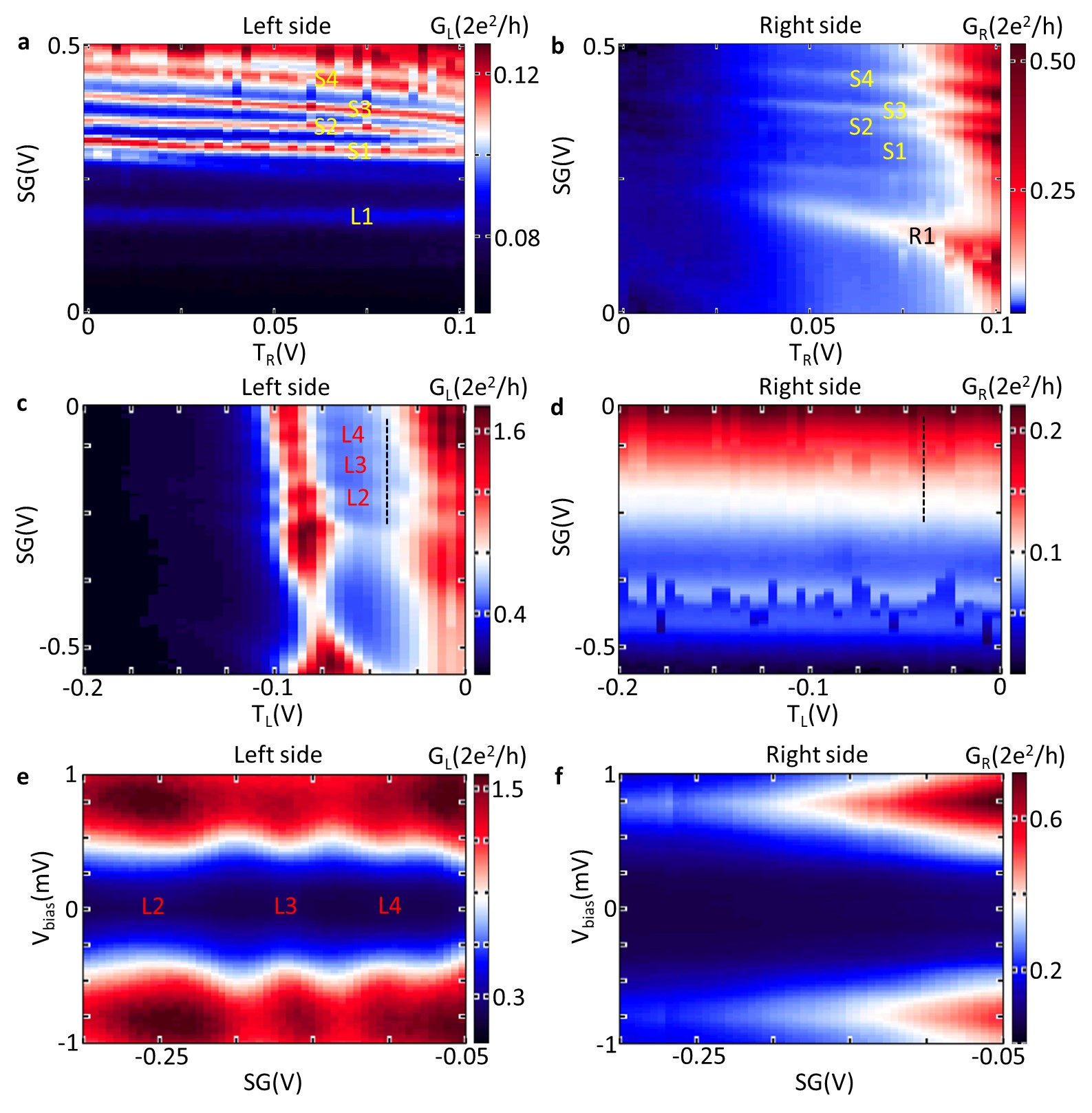}
  \caption{\textbf{Delocalized states vs. localized states in different S-gate regimes.} 
  \textbf{a} and \textbf{b}, Zero bias differential conductance $G_L$ and $G_R$ as functions of $T_R$ voltage and S-gate voltage (0 V to 0.5 V) at zero magnetic field, while $T_L$ is set to -0.13 V. 
  On the left side (panel \textbf{a}), delocalized states S1-S4 exhibit $G_R$ dependence, manifesting their non-local property. 
  On the contrary, the localized state L1 shows no $G_R$ dependence as it does not change with varying $G_R$ voltage. 
  On the right side (panel \textbf{b}), delocalzied states S1-S4 appear at the same positions with the same gate dependence as on the left side, while their magnitudes are different on the two sides. 
  Localized state L1 is missing on the right side, which is reasonable as this state is localized states near the left side. 
  Similarly, the localized state R1 only appears on the right side.
  \textbf{c} and \textbf{d}, Zero bias differential conductance $G_L$ and $G_R$ as functions of $T_L$ voltage and S-gate voltage(-0.5 V to 0 V) at B = 0 T, while $T_R$ is set to -0.15 V.
  This is the regime where we find ZBCPs close to quantization on the left side in Fig. \ref{fig2}. 
  While there are three apparent resonances(labeled as L2, L3, L4) on the left side along the black dashed line, no similar features are observed on the right side. 
  These scans confirm the low probability of having well separated Majorana bound states in that region, given the variety of localized and uncorrelated states within the nanowire. 
  \textbf{e} and \textbf{f}, Source-drain voltage vs. S-gate scans along the black dashed line in panel \textbf{c} showing the resonances on the left side and the absence of similar features on the right side.}
  \label{figs4}
\end{figure}

\begin{figure}[t!]
\centering
  \includegraphics[scale=0.25]{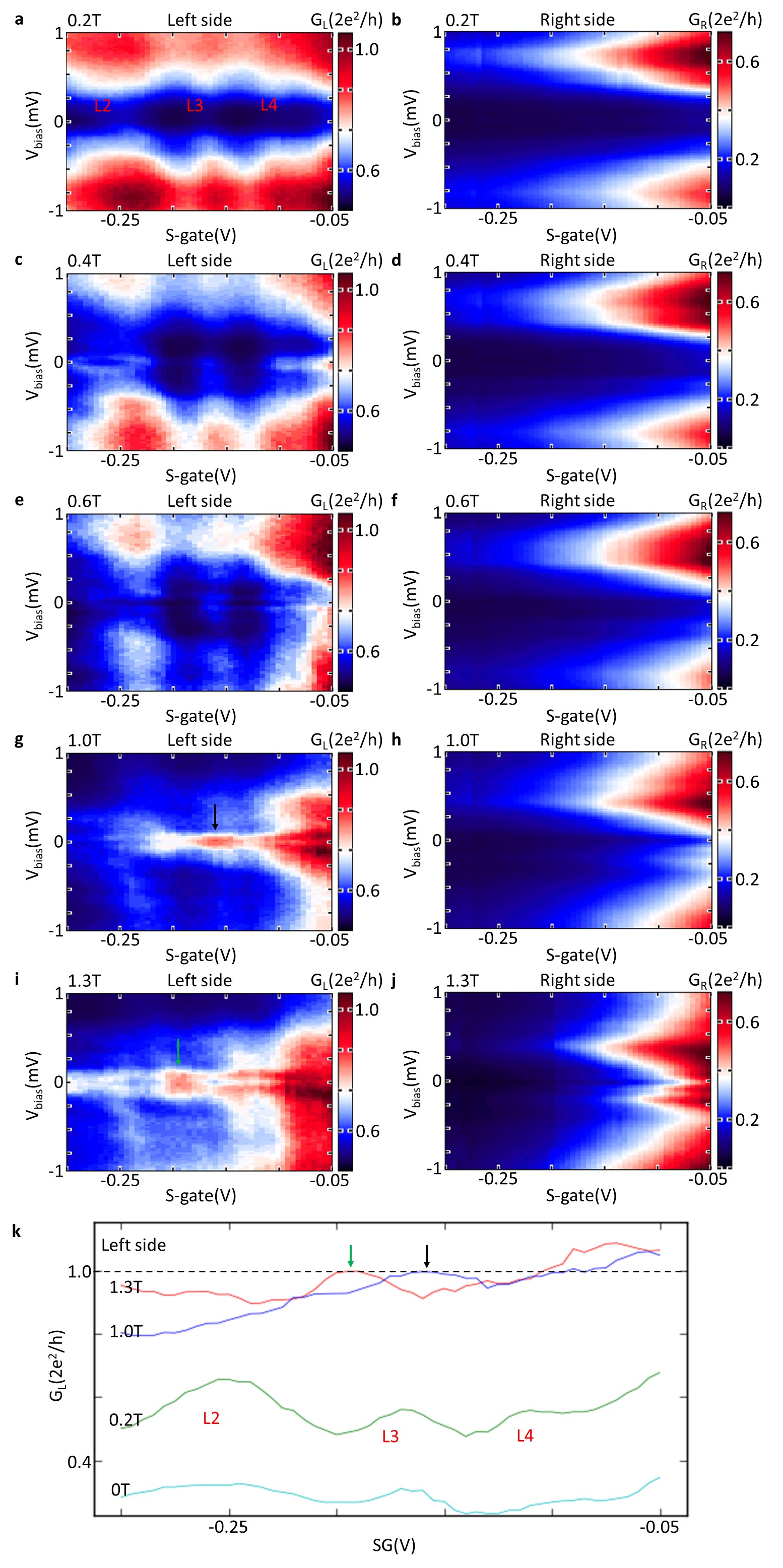}
  \caption{\textbf{Evolution of the ZBCP in magnetic fields.} 
  \textbf{a-j}, Source-drain voltage vs. S-gate scans of the same regime of Fig \ref{figs4}(c)(d) at different fields. The gate settings are $T_L$ = -0.045 V and $T_R$ = -0.105 V. 
  On the left side, subgap states and ZBCPs appear around B = 0.3 T. 
  The height of the ZBCPs reaches $2 e^{2} / h$ at 1 T (panel \textbf{g}) and again at 1.3 T (panel \textbf{j}). 
  The contact resistance of 4 $\mathrm{k} \Omega$ is subtracted for the left side. 
  On the right side, sub-gap states develop at higher fields. 
  Most importantly, no ZBCP is observed on the right side within the field range investigated. 
  \textbf{k}, Zero bias linecuts taken from Fig. \ref{figs4}(c), and panel \textbf{a}, \textbf{g}, \textbf{i} show conductance increase with increasing magnetic field and reach $2 e^{2} / h$ at 1 T and 1.3 T. }
  \label{figs5}
\end{figure}

\begin{figure}[t!]
\centering
  \includegraphics[scale=0.35]{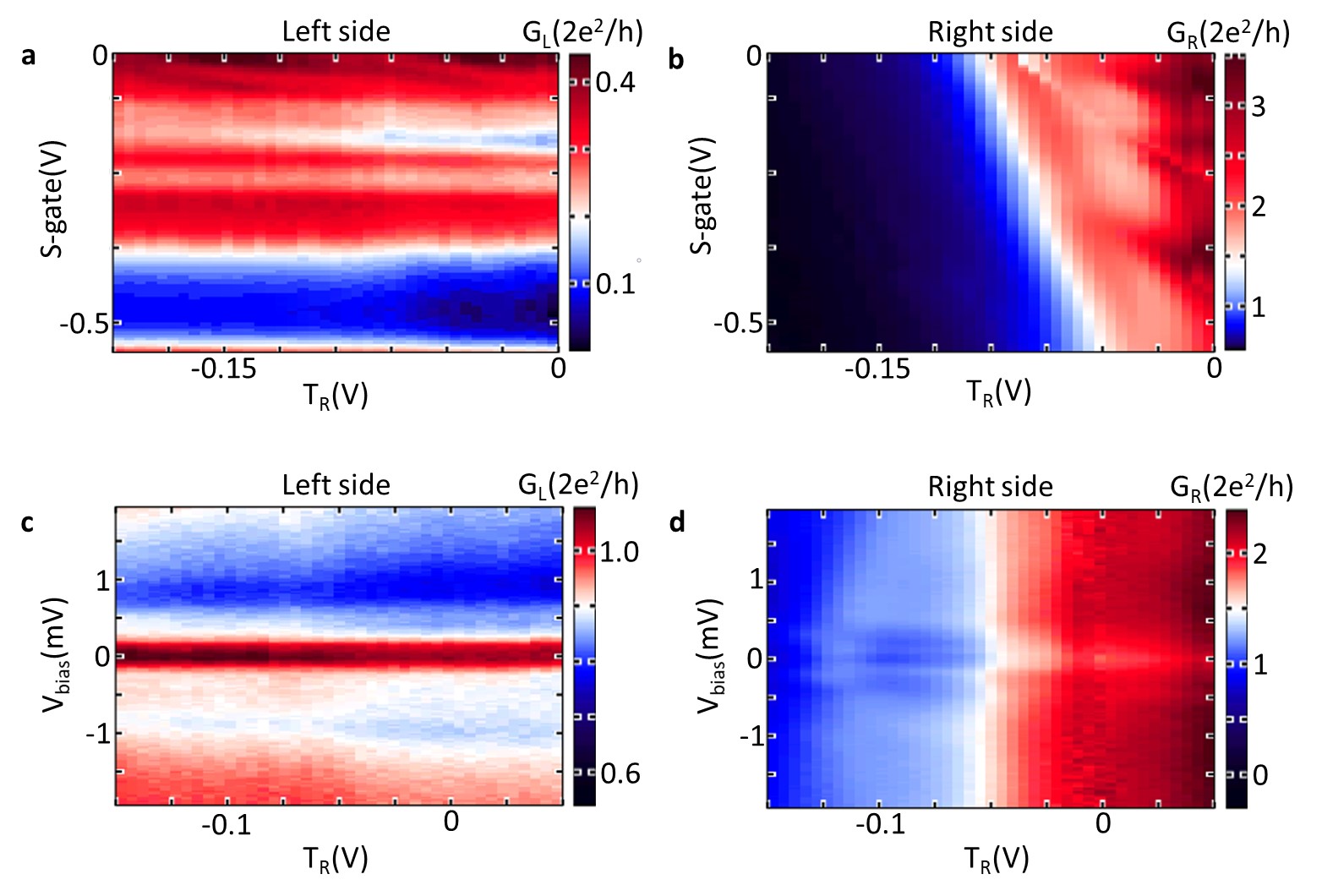}
  \caption{\textbf{Effect of $T_R$ on the left-side-only ZBCP in Fig. \ref{fig3}.} 
  \textbf{a} and \textbf{b}, Differential conductance $G_L$ and $G_R$ as functions of $T_R$ voltage and S-gate voltage at zero field and zero bias when $T_L$ is set to -0.04 V. 
  This is the regime where we find the nearly quantized  on the left side in Fig. \ref{fig2}. 
  The two sides show distinct states, which confirms the finding that there are only localized states in this regime. 
  \textbf{c} and \textbf{d},  Differential conductance $G_L$ and $G_R$ as functions of source-drain voltage and $T_R$ voltage at 1 T. 
  While the $T_R$ pinch off the right side, the ZBCP on the left side remains unchanged with conductance close to $2 e^{2} / h$. 
  Notably, there are also states near zero bias on the right side when $T_R$ is below -0.05 V. However, they never form a ZBCP. }
  \label{figs6}
\end{figure}
\clearpage

\begin{figure}[t!]
\centering
  \includegraphics[scale=0.35]{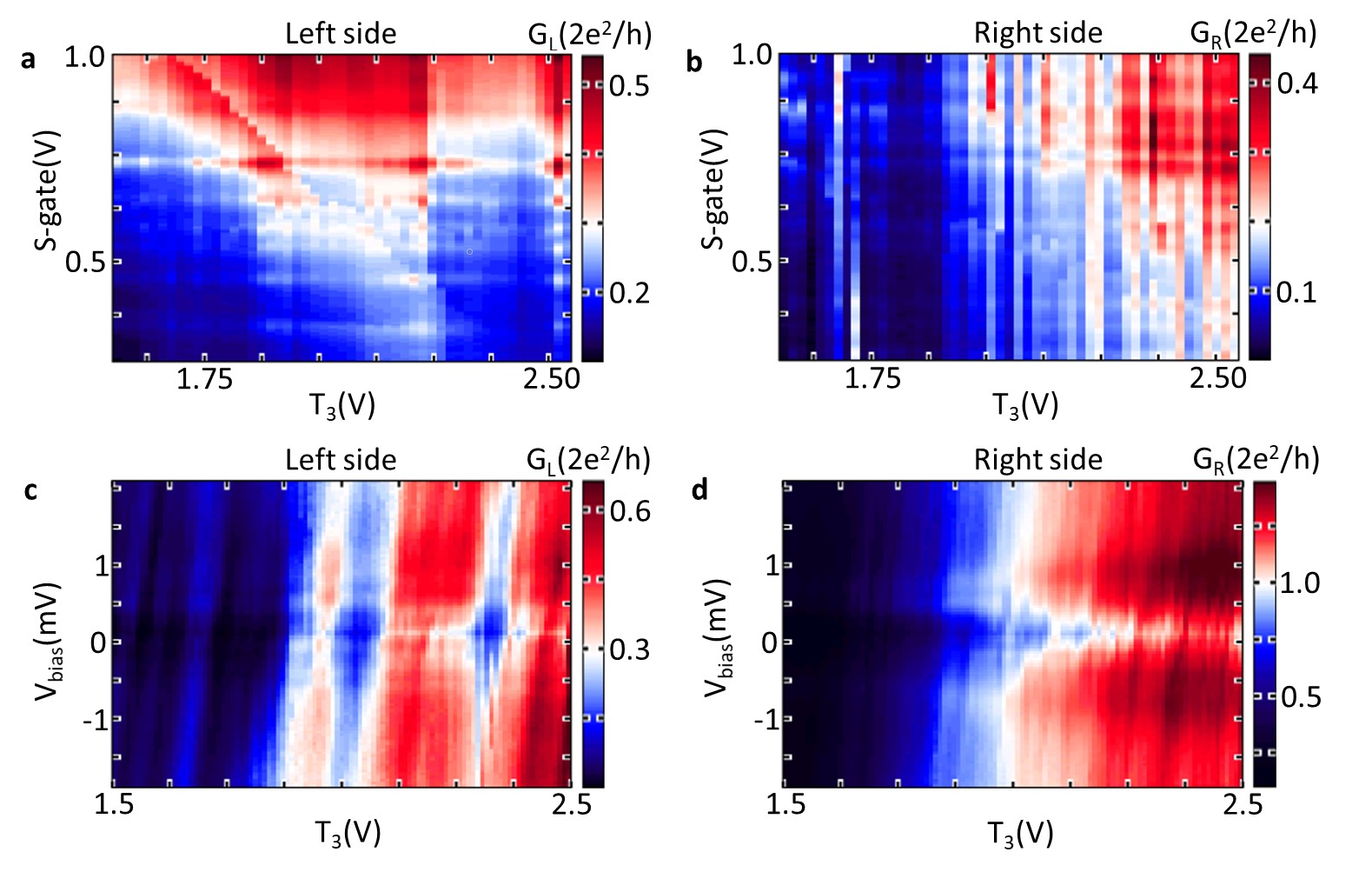}
  \caption{\textbf{Effect of gate $T_3$.} 
  The two wider barrier gates are connected and controlled by a single voltage $T_3$. 
  For all other measurements in this paper, $T_3$ is set to above 1.5 V to facilitate high transparency. 
  \textbf{a} and \textbf{b}, Differential conductance $G_L$ and $G_R$ as functions of S-gate voltage and $T_3$ voltage at zero bias and zero magnetic field, while S-gate = 1 V, $T_L$ = -0.15 V and $T_R$ = 0.1 V. 
  The resonances we observed in S-gate scans are independent of $T_3$, indicating the associated wavefunctions live far away from $T_3$. 
  $T_3$ also tune different sets of resonances on the left and the right side, which can also be seen in the source-drain voltage vs. $T_3$ scans (panel \textbf{c} and \textbf{d}). The gate settings are S-gate = 1 V, $T_L$ = -0.1 V and $T_R$ = 0.075 V. 
  Those states are independent of S-gate, indicating the existence of more dots above $T_3$.}
  \label{figs7}
\end{figure}

\begin{figure}[t!]
\centering
  \includegraphics[scale=0.35]{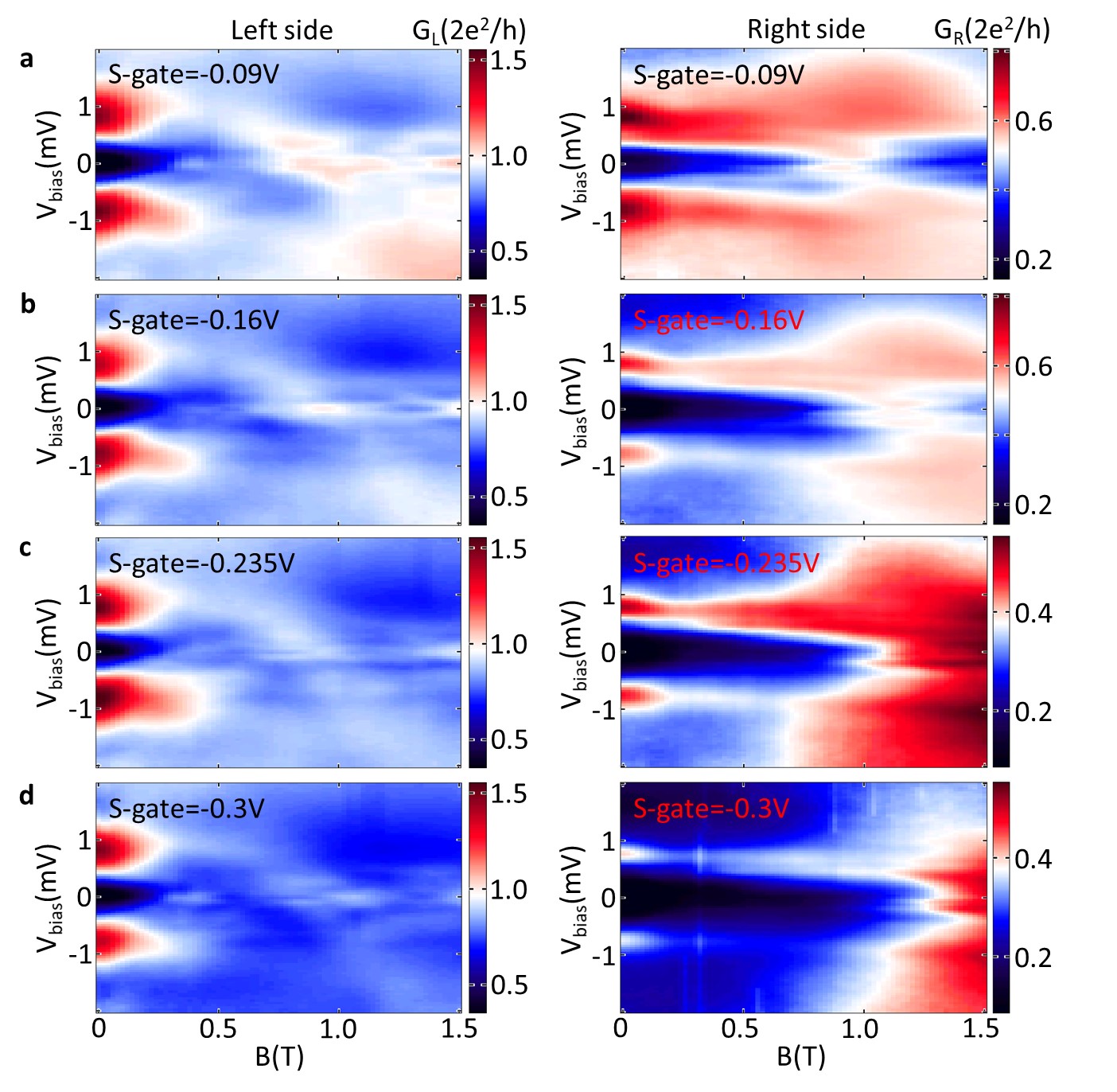}
  \caption{\textbf{Magnetic field dependence of the subgap states for different S-gate voltages}. 
  Apart from the nearly quantized ZBCP at S-gate = -0.17 V (Fig. \ref{fig2}(a)), we also found similar ZBCPs at S-gate = -0.09 V, S-gate = -0.16 V and S-gate = -0.3 V (left panels of \textbf{a}, \textbf{b} and \textbf{d}). Only when S-gate = -0.235 V (left panel of \textbf{c}), the ZBCP is lower than  $2 e^{2} / h$. The tunnel gate settings are $T_L$ = -0.045 V and $T_R$ = -0.105 V.
  The contact resistance of 4 $\mathrm{k} \Omega$ is subtracted for all the left side scans. 
  On the right side, no correlated ZBCP is observed at any of these S-gate voltages.}
  \label{figs8}
\end{figure}

\begin{figure}[t!]
\centering
  \includegraphics[scale=0.35]{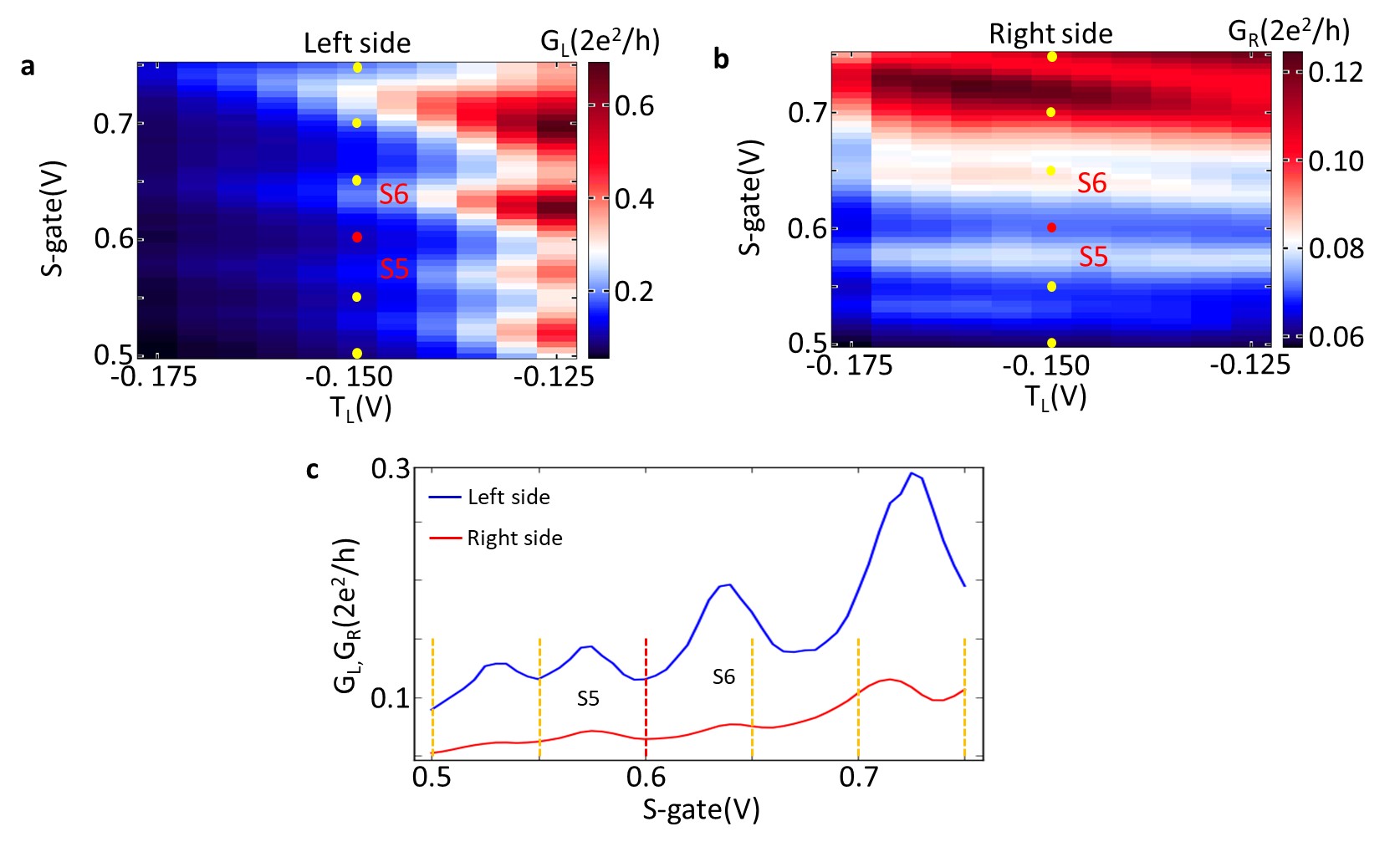}
  \caption{\textbf{Extended data for regime in Fig. \ref{fig5}.} 
  We have shown ZBCPs onset at similar fields on the left and right sides when S-gate is set to 0.6 V (Fig. \ref{fig5}). 
  As mentioned in the main text, that correlation is not robust against variations in S-gate. 
  In Figs. \ref{figs9} and \ref{figs10} we present more data around S-gate = 0.6 V. 
  \textbf{a} and \textbf{b}, differential conductance $G_L$ and $G_R$ as functions of $T_L$ voltage and S-gate voltage at zero field while bias is set to zero and $T_R$ = 0.09 V. 
  The two sides show delocalized states S5 and S6, which can be seen from the two sides simultaneously and have the same gate dependence. 
  Yellow(red) dots indicate the gate setting for Fig. \ref{figs10}(Fig. \ref{fig5}). 
  \textbf{c}, S-gate linecuts from \textbf{a} and \textbf{b} at $T_L$ = -0.15 V. 
  Yellow(red) dashed lines indicate the S-gate setting for Fig. \ref{figs10} (Fig. \ref{fig5}).}
  \label{figs9}
\end{figure}
\clearpage

\begin{figure}[t!]
\centering
  \includegraphics[scale=0.35]{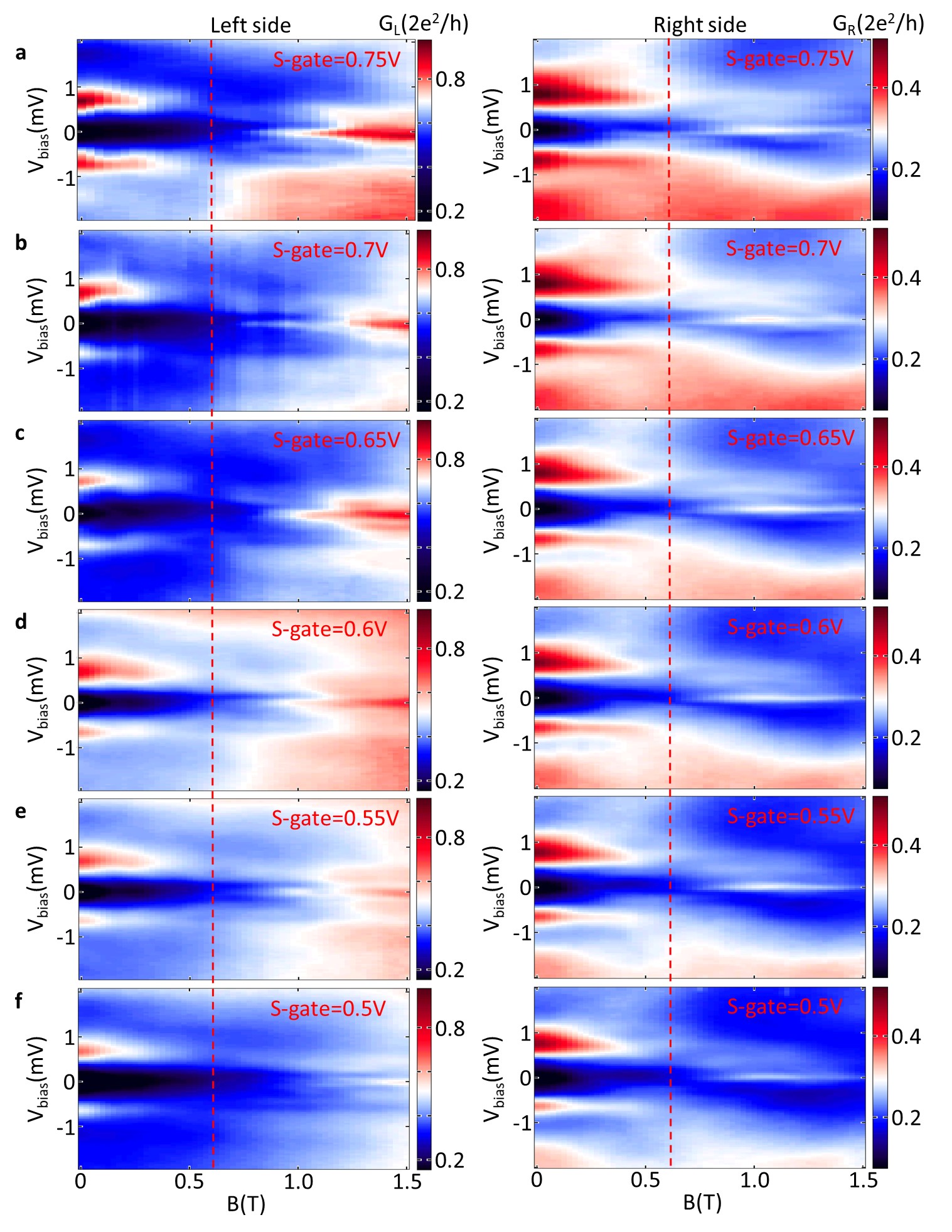}
  \caption{\textbf{Extended data for ZBCPs in Fig. \ref{fig5}.} 
  Here we present more magnetic field scans around the regime of Fig. \ref{fig5} when $T_L$ = -0.15 V and $T_R$ = 0.09 V. 
  The S-gate settings are indicated by the yellow and red dots in Fig. \ref{figs9}. 
  As shown in the left panels, the onset fields of the ZBCPs on the left side change to higher fields when S-gate is reduced. 
  And the ZBCPs also exhibits splitting features at S-gate = 0.55 V and S-gate = 0.5 V. 
  On the right side, however, the ZBCP onset and splitting do not generally match the left side manifestations.}
  \label{figs10}
\end{figure}

\begin{figure}[t!]
\centering
  \includegraphics[scale=0.35]{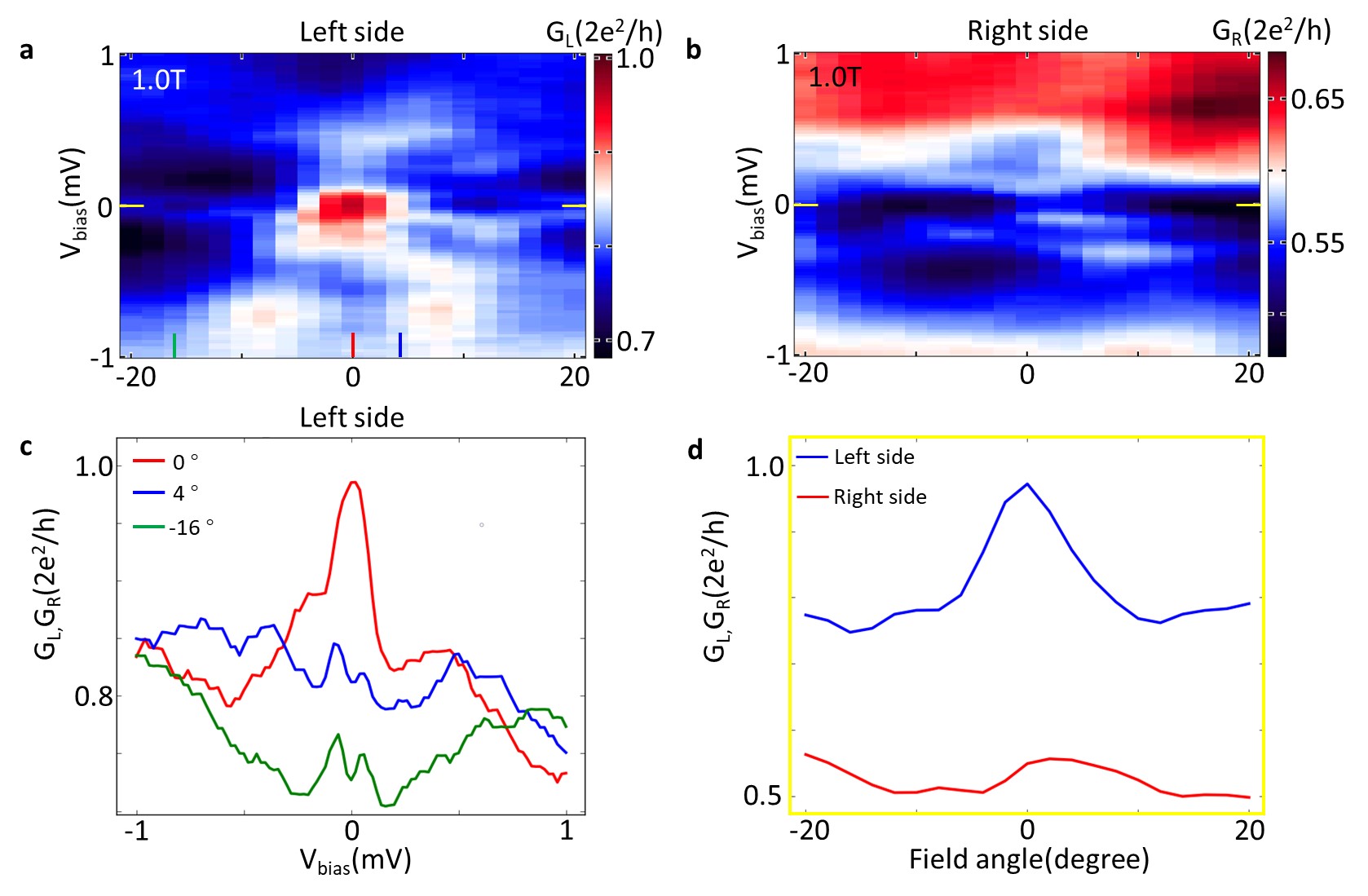}
  \caption{\textbf{Magnetic field angle dependence of the ZBCP and subgap states on both sides. 
  a} and \textbf{b}, Differential conductance $G_L$ and $G_R$ as functions of field angle and source-drain voltage when $T_R$ = -0.105 V, $T_L$ = -0.04 V, S-gate = -0.18 V. 
  Note the contact resistance of 4 $\mathrm{k} \Omega$ is subtracted for the left side. 
  The field is parallel to the nanowire and perpendicular to the spin-orbit field when the field angle is zero. 
  On the left side, the ZBCP only exists and reaches $2 e^{2} / h$ within a small angle around zero degree. 
  On the right side, the subgap states are asymmetrical in field angle. 
  Most importantly, no ZBCP is observed in the range -20 degree to 20 degree. 
  \textbf{c}, Bias linecuts at 0, 4, -16 degree field angle from panel \textbf{a}. 
  The ZBCP splits into two peaks when the field angle deviates from 0 degree. 
  \textbf{d}, Zero bias linecuts show distinct behavior on the two sides: the zero bias conductance on the left side peaks at zero degree while the zero bias conductance on the right side remains almost unchanged}
  \label{figs11}
\end{figure}

\begin{figure}[t!]
\centering
  \includegraphics[scale=0.35]{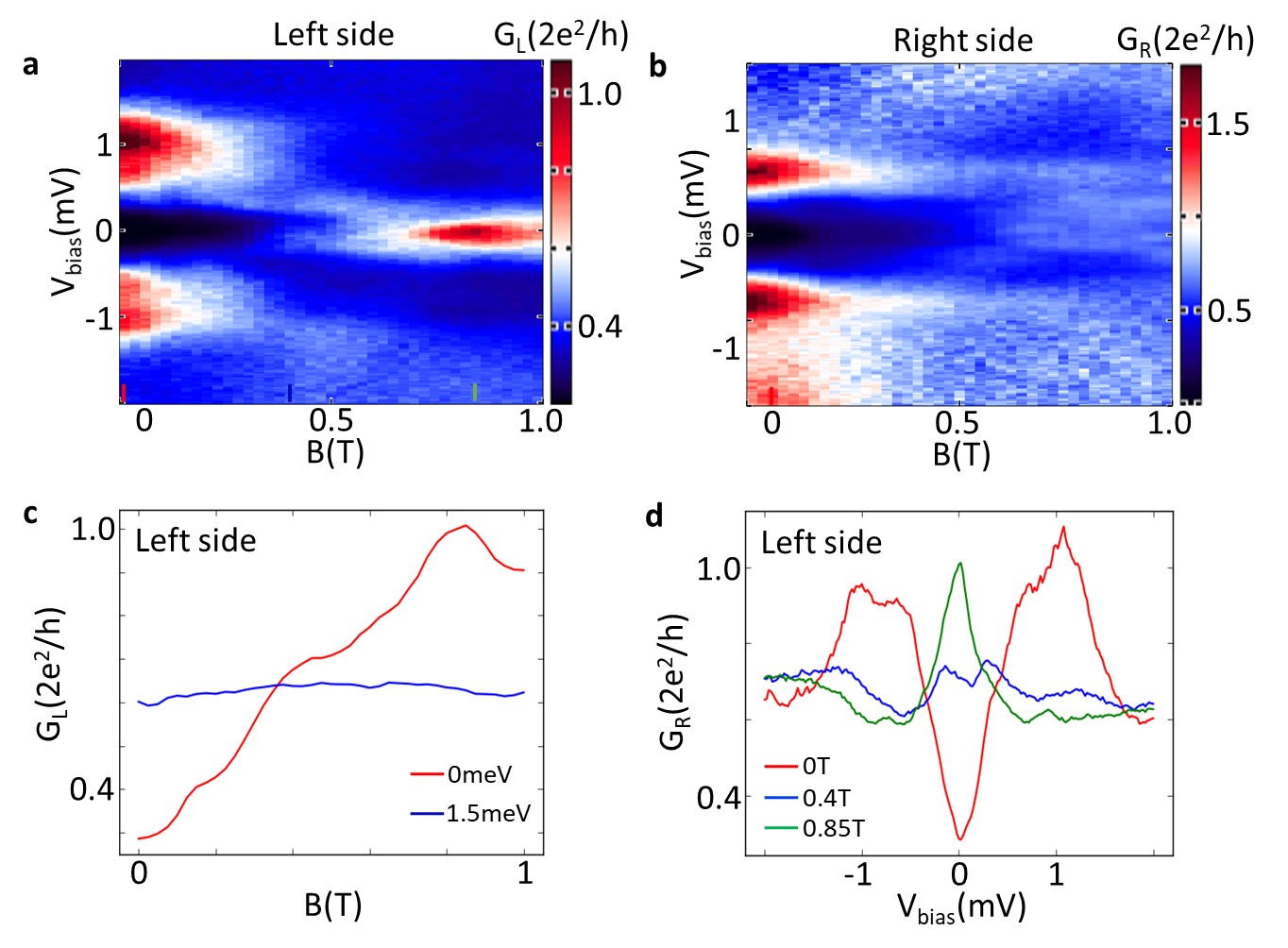}
  \caption{\textbf{Nearly quantized ZBCP in the more open S-gate regime.} 
  \textbf{a} and \textbf{b}, Differential conductance $G_L$ and $G_R$ as functions of bias voltage and magnetic field when S-gate = 1 V, $T_L$ = -0.25 V and $T_R$ = 0.05 V. This is a regime with a more positive S-gate setting than that explored in Figs. 1-3. 
  On the left side, a ZBCP appears around B = 0.6 T and reaches $2 e^{2} / h$ around 0.85 T. 
  The contact resistance of 4$\mathrm{k} \Omega$ is subtracted for the left side again. 
  While there are some subgap states on the right side around similar fields, no prominent ZBCP is observed. 
  \textbf{c}, Magnetic field linecuts at $V_{bias}$ = 0 and $V_{bias}$ = 1.5 meV from panel \textbf{a}. 
  \textbf{d}, Bias linecuts at $B$ = 0, 0.4 T and 0.85 T from panel \textbf{a}.}
  \label{figs12}
\end{figure}
\clearpage

\end{widetext}
\end{document}